\documentclass[12pt]{article}
\usepackage{amsmath,amsfonts, amssymb}
\usepackage{graphicx}
\usepackage{longtable}
\usepackage{url}
\usepackage{tocloft}
\makeatletter
 
  \@addtoreset{equation}{section}
\makeatother
\def\beqar {\begin{eqnarray}}
\def\eeqar {\end{eqnarray}}
\def\beq {\begin{equation}}
\def\eeq {\end{equation}}
\def\A{{\cal A}}

\def\al{\alpha}
\def\bt{\beta}
\def\del{\delta}

\def\ga{\gamma}
\def\Ga{\Gamma}

\def\ep{\epsilon}
\def\la{\lambda}
\def\La{\Lambda}
\def\om{\omega}
\def\Om{\Omega}
\def\th{\theta}

\def\d{\partial}
\def\bd{{\bar \partial}}

\def\ba{{\bar a}}
\def\bz{{\bar z}}

\def\bom{{\bar \omega}}

\def\hf{\frac{1}{2}}

\def\<{\langle}

\def\>{\rangle}

\def\re{{\rm Re}}
\def\im{{\rm Im}}

\def\dim{{\rm dim}}

\def\C{\mathbb{C}}
\def\Q{\mathbb{Q}}
\def\R{\mathbb{R}}
\def\Z{\mathbb{Z}}
\def\F{\mathbb{F}}
\def\H{\mathbb{H}}

\begin{document}

\begin{titlepage}
\null\vspace{-62pt} \pagestyle{empty}
\begin{center}
\vspace{0.8truein}

{\Large\bf
Abelian Chern-Simons theory on the torus \\
\vspace{.35cm}
and physical views on the Hecke operators
} \\

\vspace{1.0in} {\sc Yasuhiro Abe} \\
\vskip .12in {\it Cereja Technology Co., Ltd.\\
3-11-15 UEDA-Bldg. 4F, Iidabashi   \\
Chiyoda-ku, Tokyo 102-0072, Japan }\\
\vskip .07in {\tt abe@cereja.co.jp}\\
\vspace{1.3in}

\centerline{\large\bf Abstract}
\end{center}

\noindent
In the previous paper arXiv:1711.07122, we show that
a holomorphic zero-mode wave function in
abelian Chern-Simons theory on the torus can be considered
as a quantum version of a modular form of weight 2.
Motivated by this result, in this paper
we consider an action of a Hecke operator on
such a wave function from a gauge theoretic perspective.
This leads us to obtain some physical views on the
Hecke operators in number theory.

\end{titlepage}
\pagestyle{plain} \setcounter{page}{2} 

\tableofcontents
\vskip 1.2cm


\section{Introduction}

It has been known for a long time that
holomorphic part of zero-mode wave functions in abelian Chern-Simons (CS)
theory on the torus can be expressed in terms of a Jacobi theta function
in the context of geometric quantization \cite{Bos:1989wa,NairBook,Nair:2016ufy}.
(For foundations of the geometric quantization, see, {\it e.g.},
\cite{Woodhouse:1992de,Blau:1992,Ali:2004ft}.)
Since any complex functions on the torus can, by definition, be expressed
in terms of elliptic functions, this result sounds natural.
Indeed it is well-known that the so-called Jacobi elliptic  functions can
be defined in terms of the Jacobi theta functions.
Strictly speaking, however, the resultant form of the holomorphic wave function is
not invariant under doubly periodic translations.
This implies that we can not make a smooth transition from
classical functions on the torus to quantum wave functions on
the same manifold.
One may interpret this matter as a reflection of
ambiguities buried in a quantization process.
We would, however, expect to make a quantization such that
the doubly periodicity also holds in the quantum wave function.
In fact, such a quantization has been reported previously by the author
\cite{Abe:2008wn,Abe:2010sa}.
One of the main purposes of this paper is to deriver
a detailed review of this quantization procedure.

In order to clarify the issue, we now state some technical aspects
of the situation as follows.
In the geometric quantization the transition from a classical theory
to a quantum counterpart is realized by imposing a polarization
condition on a prequantum wave function parametrized by canonical
coordinates on a symplectic manifold.
The polarization condition can straightforwardly be implemented
by use of holomorphicity if
the symplectic manifold also holds complex structure, that is,
the manifold is K\"ahler.
This is relevant to the present case since the tours is a K\"ahler manifold.
The holomorphic coordinates $(z ,\bz )$ of the tours are
therefore a suitable choice to parametrize wave function.
This choice is, however, not compatible with
a classical picture of the doubly periodic functions
which are basically parametrized by $\re z$ and $\im z$.
This is a main reason for the above-mentioned discrepancy
between classical and quantum wave functions on the torus.
The discrepancy may be solved if we describe the prequantum wave
function in terms of $\re z$ and $\im z$. Upon the quantization, however,
we need to introduce the holomorphic coordinates $(z ,\bz )$
otherwise we can not suitably impose the polarization condition.
We thus need to define the wave function in accord with these requirements.
This is exactly what we shall carry out in section 3 of the present paper.

This paper is also motivated by an interest in applications of quantum field
theory to number theory.
According to the modularity theorem,
or the formerly-called Taniyama-Shimura-Weil conjecture, an elliptic curve
over rational number $\Q$ and a modular form of weight 2 are in one-to-one
correspondence at the level of $L$-functions.
On the other hand, the elliptic curves, extended to the field
of complex number $\C$, can be described by the elliptic functions,
as typically represented by the Weierstrass $\wp$ function.
As mentioned above the very elliptic functions can be considered
as complex functions on the torus.
These interrelations at least suggest a possibility to connect
the holomorphic zero-mode wave functions in abelian CS theory on the torus with
the modular forms weight 2.
For introduction to the modular forms and related subjects such as elliptic curves
and $L$-functions, see, {\it e.g.},
\cite{Koblitz:1993bk,Ono:2004bk,Kurokawa:2005bk,Stein:2007bk,Lozano-Robledo:2011bk}.
The online database \cite{LMFDB} on the $L$-functions and the modular forms is also useful.

Partly motivated by these thoughts, in the previous paper \cite{Abe:2017mf}
we argue that the holomorphic zero-mode wave function can quantum theoretically
considered as the modular form of weight 2.
Another purpose of the present paper arises from a natural extension
of this result, that is, we like to make use of this result so as
to find a physical perspective on a problem in number theory.
Particularly, we are interested in a physical interpretations
of a Hecke operator acting on the modular forms.
In the literature, physical views on the Hecke operator
have been considered previously.
In \cite{Rajeev:2002xd} it is discussed that finding simultaneous eigenvectors
of the Hecke operator is analogous to determining simultaneous eigenfunctions
of a hermitian operator in quantum mechanical systems;
in particular, connection between eigenvalues of the Hecke operators and spectra of
hermitian random matrices at a certain limit has been suggested.
The Hecke operators also appear
in the study of elliptic genera in superconformal field theories
\cite{Dijkgraaf:1996xw,Gukov:2004fh} as well as in the context of
the geometric Langlands program \cite{Kapustin:2006pk}.
These topics are beyond the scope of this paper;
we here simply focus on an interpretation of the Hecke operator
as a hermitian operator acting on the holomorphic zero-mode wave function
in abelian CS theory on the torus.

The organization of this paper is as follows.
In the next section we review the geometric quantization
of abelian CS theory on the torus, following Nair's formulation
\cite{NairBook,Nair:2016ufy}.
In section 3, as mentioned above, we show that the holomorphic
zero-mode wave function can obey the doubly periodic condition
when the level number of the abelian CS theory is even.
We derive this relation by imposing gauge invariance on
the zero-mode wave function where the gauge transformations
are induced by the doubly periodic translations
of the zero-mode variable.
In section 4 we review main results in the previous paper \cite{Abe:2017mf}.
In section 5 we briefly introduce some basic facts on the Hecke operators and
the $L$-functions for the modular forms,
including related topics such as level $N$ congruence subgroups of the modular group
and corresponding $L$-functions.
We then consider how the Hecke operators act on the holomorphic wave function
constructed in section 3.
We argue that the action of the Hecke operator can be interpreted
as a sum over the above-mentioned gauge transformations of the holomorphic wave function.
This automatically explains that the holomorphic wave function
is an eigenform of the Hecke operator.
We argue that the notion of the level which is inherent in
the modular forms also appears in the holomorphic wave function.
Other speculative physical views on the Hecke operators are also
discussed in this section.
Lastly, in section 6 we present brief conclusions.

\section{Geometric quantization of abelian CS theory on the torus}

In this section we briefly review geometric quantization of
abelian Chern-Simons (CS) theory on the torus,
following \cite{Bos:1989wa,NairBook,Nair:2016ufy}.
The geometric quantization is carried out on the zero-mode
part of the abelian CS gauge field.
A guiding concept of the quantization is a K\"ahler form of
the torus on which the zero-mode variable is defined.
In the following we see that all the key ingredients of
the geometric quantization, such as a K\"ahler potential, a symplectic
potential and a zero-mode wave function, can be derived
from the K\"ahler form.


The torus can be described in terms of two real coordinates $\xi_1$, $\xi_2$,
satisfying the periodicity condition $\xi_r \rightarrow \xi_r + {\rm (integer)}$
where $r=1,2$.
In other words, $\xi_r$ take real values in $0 \le \xi_r \le 1$,
with the boundary values $0$, $1$ being identical.
Complex coordinates of the torus can be
parametrized as $z = \xi_1 + \tau \xi_2$ where $\tau \in \C$ is
the modular parameter of the torus.
By definition, we can impose the doubly periodic condition on $z$.
Namely, functions of $z$ are invariant under the doubly periodic translations
\beq
    z \, \rightarrow \, z + m + n \tau
    \label{2-1}
\eeq
where $m$ and $n$ are integers.
Notice that we can absorb the real part of $\tau$ into $\xi_1$
without losing generality.
In the following, we then assume $\re \tau = 0$, {\it i.e.},
\beq
    \tau \, = \, \re \tau + i \im \tau \, =  \, i \im \tau  \, := \, i \tau_2
    \label{2-2}
\eeq
with $\tau_2 > 0$.

The torus has a holomorphic one-form $\om = \om (z) dz $, satisfying
\beq
    \int_\al \om = 1 \, , ~~~ \int_\bt \om = \tau = i \tau_2
    \label{2-3}
\eeq
where the integrals are made along two non-contractible cycles
on the tours, which are conventionally labeled as $\al$ and $\bt$ cycles.
The one-form $\om$ is a zero mode of the anti-holomorphic derivative
$\d_\bz = \frac{\d}{\d \bz}$. We can assume $\om (z) = 1$.
In terms of $\om$ the gauge potential of CS theory on the torus
can be parametrized as
\beq
    A_\bz  =   \d_\bz \th +  \frac{ \pi \bom }{ \tau_2 } a
    \label{2-4}
\eeq
where $\th$ is a complex function $\th (z , \bz )$ and
$a$ is a complex number corresponding to the value of
$A_\bz$ along the zero mode of $\d_z$.
The abelian gauge transformations can be represented by
\beq
    \th \, \rightarrow \, \th + \chi
    \label{2-5}
\eeq
where $\chi$ is a complex constant or a phase factor of the $U(1)$ theory.
With a suitable choice of $\chi$ we can parametrize the
gauge potential solely by the zero-mode contributions, $a$ and its complex
conjugate $\ba$:
\beq
    A_{z} \, = \,   \frac{\pi \om}{ \tau_2 } \ba \, , ~~~~
    A_{\bz} \, = \,  \frac{\pi \bom}{\tau_2 } a \, .
    \label{2-6}
\eeq
Since the complex variable $a$ is defined on the torus it is natural to
require that physical observables of the zero modes are invariant under
the doubly periodic translations
\beq
    a \, \rightarrow \, a + m + i n \tau_2
    \label{2-7}
\eeq
where $m$ and $n$ correspond to the
winding numbers along the $\al$ and $\bt$ cycles, respectively.
It is known that the complex torus
can be embedded into a complex projective space, that is, the zero-mode
variable $a$ may satisfy the scale invariance
\beq
    a \, \sim \, \la \, a
    \label{2-9}
\eeq
where $\la$ is a complex constant.

\vskip 0.5cm \noindent
\underline{Geometric quantization as a K\"ahler-form program}

All the important ingredients
in geometric quantization of abelian Chern-Simons theory on the torus
are derived from a K\"ahler form of the torus parametrized by
the zero-mode variable $a$. From (\ref{2-6}) the zero-mode K\"ahler form
is defined as
\beq
    \Om^{(\tau_2 )} \, = \, \frac{l}{2 \pi} da \wedge d\ba \int_{z,\bz}
    \frac{ \pi \bom}{\tau_2}    \wedge  \frac{ \pi \om}{\tau_2}
    \, = \, i \frac{\pi l}{ \tau_2} da \wedge d \ba
    \label{2-10}
\eeq
where the integral is taken over $dz d \bz$ and
$l$ is the level number associated to the abelian Chern-Simons theory.
We here use the normalization of $\om$ and $\bom$ given by
\beq
    \int_{z,\bz} \bom \wedge \om  \, = \, i 2  \tau_2 \, .
    \label{2-11}
\eeq
A K\"ahler potential $K( a, \ba)$ associated with the zero-mode K\"ahler form
$\Om^{(\tau_2 )}$ is defined as
\beq
    \Om^{(\tau_2 )} \, =  \,   i \d \bd K (a , \ba )
    \label{2-12}
\eeq
where $\d$, $\bd$ denote the Dolbeault operators.
The is definition leads to
\beq
    K (a, \ba ) \, = \, \frac{\pi l}{ \tau_2 } a \ba + u (a ) + v ( \ba )
    \label{2-13}
\eeq
where $u(a)$ and $v( \ba )$ are purely holomorphic and
anti-holomorphic functions, respectively. These functions
represent {\it ambiguities} in the choice of $K (a, \ba )$.

A symplectic potential (or a canonical one-form) $\A^{(\tau_2 )}$
corresponding to the K\"ahler form $\Om^{(\tau_2 )}$ is defined as
\beq
    \Om^{(\tau_2 )} \, =  \, d \A^{(\tau_2 )} \, .
    \label{2-14}
\eeq

In the program of geometric quantization a {\it quantum} wave function
$\Psi [ A_\bz ]$ generally satisfies the  so-called polarization condition
\beq
    \left( \d_\ba + \frac{1}{2} \d_\ba K \right) \, \Psi [ A_\bz ] \, = \, 0
    \label{2-17}
\eeq
where $K = K ( a , \ba )$ is the zero-mode K\"{a}hler potential in (\ref{2-13}).
The polarization condition leads to the specific form
\beq
    \Psi[ A_\bz ] \, = \, e^{-\frac{K}{2}} \psi[ A_\bz ]
    \label{2-18}
\eeq
where $\psi [ A_\bz ]$ is a holomorphic function of $A_\bz$.
In the present case the physical variables are given by $a$ and $\ba$
so that the wave function can be expressed as
\beq
    \Psi[ A_\bz] \, := \, \Psi[a , \ba ] \, = \, e^{-\frac{K (a,\ba)}{2}} f(a)
    \label{2-19}
\eeq
where $f(a)$ is a function of $a$. We call $f(a)$ a {\it holomorphic zero-mode
wave function}.
Notice that we here define $\Psi[ a , \ba ]$ with $K ( a , \ba )$ including
the above-mentioned ambiguities in its choice.


An inner product of the zero-mode wave functions $\Psi[a , \ba ]$ in (\ref{2-19})
can be expressed as
\beqar
    \< \Psi | \Psi^\prime \> & = &
    \int d\mu(a,\ba)  \, \overline{\Psi[ A_\bz ]} \Psi^\prime [ A_\bz ]
    \nonumber \\
    & = &
    \int d\mu(a,\ba) \, e^{-K(a,\ba)} \, \overline{f(a)} f^\prime (a)
    \label{2-22}
\eeqar
where $\overline{f(a)}$ is the complex conjugate of $f (a)$ and
$d \mu (a , \ba )$ denotes the integral measure of the zero-mode
variable on the torus. The meaning of the integral is therefore the
same as those in (\ref{2-10}) and (\ref{2-11}).

\section{Double periodicity in holomorphic wave functions}

Since the zero-mode variable $a$ is defined on the torus, the holomorphic
zero-mode wave function
$f(a)$ is expected to be invariant under the doubly periodic
translations $a \rightarrow a + m + i n \tau_2$ ($m , n \in \Z$).
Indeed, as shown in \cite{Abe:2008wn,Abe:2010sa},
we can argue that such an expectation is true for $l \in 2 \Z $.
To be more specific, a gauge invariance condition
on the zero-mode wave function $\Psi [a , \ba ]$
with certain choices of $\A^{(\tau_2 )}$ and $K$
(where the gauge transformations are induced by the doubly periodic
translations) leads to the relation
$f(a + m + i n \tau_2 ) = e^{i \pi l mn } f(a)$.
In this section we shall carry a careful review
of this relation.

\vskip 0.5cm \noindent
\underline{Change of variables and the symplectic structure of the torus}

From our setting $\tau = i \tau_2$ in
(\ref{2-2}), the complex coordinate $z$ on the torus is written as
$z = \xi_1 + i \tau_2 \xi_2$.
It is then useful to choose coordinates on the torus as
$z_1 := \bz - z = - 2i \tau_2 \xi_2$
and $z_2  := \tau \bz - \bar{\tau} z = 2i \tau_2 \xi_1$
so that we can parametrize the torus in terms of
the real variables along the non-contractible cycles.
This parametrization does not mean that the torus
loses the complex structure, of course.
It is, however, useful to connect the coordinates
directly to the doubly periodic transformations in (\ref{2-1}).
In fact, this parametrization corresponds to a conventional
definition of a double periodic complex function;
$f(x, y ) = f (x + m , y + n)$ where $z = x + i y$ and $m , n \in \Z$.

We then reparametrize the complex zero-mode variables $( a , \ba )$ by
\beqar
    a_1 &= & \ba - a \, ,
    \label{3-1} \\
    a_2 &=& \tau \ba - \bar{\tau} a \, = \, i \tau_2 ( \ba + a) \, .
    \label{3-2}
\eeqar
Notice that $a_1 = - i 2 (\im \, a)$, $a_2 =  i 2 \tau_2 ( \re \, a)$.
These variables
essentially represent the real and imaginary parts of the complex variable $a$.
In terms of these the integral measure $d \mu (a , \ba)$
may be expressed as $d ( \re \, a) d (\im \, a)$.
On the other hand, the complex structure of the torus and, hence, a holomorphic
quantization program should be described in terms of $(a, \ba )$.
Thus, for the sake of the geometric quantization, {\it per se},
it is not appropriate to use the variables $( a_1 , a_2 )$.
In other words, for the construction of zero-mode wave functions
we still need to keep using the polarization condition (\ref{2-17}).
As seen in a moment, we can, however, express a symplectic two-form of the torus
in terms of $(a_1 , a_2)$.
A corresponding symplectic potential and
an analog of the K\"ahler potential can then be expressed
in terms of $(a_1 , a_2)$.  This suggests that at least at
a level of choosing the K\"ahler potential (\ref{2-13}) we can
use the canonical coordinates $(a_1 , a_2)$.
In what follows we shall clarify these points by
reviewing some results in \cite{Abe:2008wn,Abe:2010sa}.

In terms of $( a_1 , a_2 )$ the zero-mode part of
the abelian CS gauge potentials are written as
\beq
    A_{\xi_1} \, = \,   \frac{\pi \om_2}{ \tau_2 } \frac{a_1}{ \tau_2 } \, , ~~~~
    A_{\xi_2} \, = \,  \frac{\pi \om_1}{\tau_2 } \frac{a_2}{ \tau_2 }
    \label{3-3}
\eeq
where the associated one-forms $\om_1$, $\om_2$ are defined as
\beq
    \om_1 \, = \, \frac{d z_1}{ 2i}  \, = \, - \tau_2 d\xi_2  \, , ~~~~
    \om_2 \, = \, \frac{d z_2}{ 2i} \, = \,  \tau_2 d\xi_1 \, .
    \label{3-4}
\eeq
The normalization for $\om_1$ and $\om_2$ can be given by
\beq
    \int_{z,\bz} \frac{\om_1}{\tau_2} \wedge \frac{\om_2}{\tau_2} \, = \, 1 \, .
    \label{3-5}
\eeq
In terms of these the holonomies of the torus (\ref{2-3}) are simplified as
\beq
    \oint_{\al_r} \om_s \, = \,   \ep_{rs} \, \tau_2
    \label{3-6}
\eeq
where $\ep_{rs}$ ($r,s = 1,2$) denotes the rank-2 Levi-Civita symbol
and $\al_1$, $\al_2$ correspond to the $\al$ and $\bt$ cycles, respectively.

A zero-mode symplectic two-form is then expressed as
\beq
    \Om^{(\tau_2 )} \, = \,  \frac{l}{2 \pi \tau_2^2 }
    da_1 \wedge d a_2
    \int_{z,\bz} \frac{\pi \om_2}{\tau_2} \wedge \frac{ \pi \om_1}{\tau_2}
     \, = \,- \frac{ \pi l}{2 \tau_2^2 } da_1 \wedge d a_2 \, .
    \label{3-7}
\eeq
The corresponding symplectic potential (or the canonical one-form) can be written as
\beq
    \A^{(\tau_2 )} \, = \,
    \frac{\pi l}{4 \tau_2^2} \int_{z,\bz}
    \left(
    \frac{\om_2 a_1}{\tau_2} \wedge \frac{\om_1}{\tau_2} da_2
    - \frac{\om_1 a_2}{\tau_2} \wedge \frac{\om_2}{\tau_2} da_1
    \right)
    \, =  \,
    - \frac{ \pi l}{4 \tau_2^2 } \left( a_1 da_2 + a_1 d a_2 \right) \, .
    \label{3-8}
\eeq
The explicit form of the symplectic two-form $\Om^{(\tau_2 )}$ in (\ref{3-7})
means that $a_1$ and $a_2$ can be served as canonical coordinates of the torus.
Although these are not complex conjugate to each other,
it is suggestive that we can define an analog
of a K\"ahler potential corresponding to $\Om^{(\tau_2 )}$ in (\ref{3-7}):
\beq
    W ( a_1 , a_2 )  \, = \,
    i \frac{ \pi l}{2 \tau_2^2 } a_1 a_2
    \, = \, - \frac{\pi l}{2 \tau_2 }
    (  \ba^2 - a^2 ) \, .
    \label{3-9}
\eeq
This is in a form of separation of holomorphic and antiholomorphic parts.
We are thus allowed to rewrite the K\"{a}hler potential $K ( a , \ba )$ in (\ref{2-13}) as
\beq
    K ( a, \ba ) \, = \, K_0 + W + u + {\bar v}
    \label{3-10}
\eeq
where
\beq
    K_0  =  \frac{\pi l}{ \tau_2 } a \ba  \,  , ~~~ W \, = \, W (a_1 , a_2 )  \,  , ~~~
    u=u( a ) \,  , ~~~ {\bar v}=v( \ba ) \, .
    \label{3-11}
\eeq
As before, $u( a )$ and $v( \ba )$ denote holomorphic and antiholomorphic functions, respectively.

The zero-mode wave function (\ref{2-19}) is written as
\beq
    \Psi[ a, \ba ] \, = \, e^{- \frac{K_0 + u + \bar{v}}{2} } \, F [a , \ba ]
    \label{3-12}
\eeq
where
\beq
    F [ a, \ba ] \, = \, e^{ - \frac{W}{2} } \, f ( a ) \, .
    \label{3-13}
\eeq
The inner product (\ref{2-22}) is then expressed as
\beq
    \< \Psi | \Psi^\prime \> \, = \,
    \int d\mu(a,\ba) \, e^{-( K_0 + u + \bar{v} )} \, \overline{f(a)} f^\prime (a)
    \label{3-14}
\eeq
where we use $\overline{F [ a, \ba ] } = e^{ \frac{W}{2} } \overline{f(a)}$,
$F^\prime [ a, \ba ] = e^{ - \frac{W}{2} } f^\prime (a) $.
The wave function $F [a ,\ba ]$ is relevant to the choice of the symplectic
form $\Om^{(\tau_2 )}$ in (\ref{3-7}) or the choice of the canonical coordinates
$(a_1 , a_2 )$.
The factor of $e^{- \frac{W}{2}}$ in (\ref{3-13}) is then appropriate
one in the definition of $F[a , \ba]$.
We now argue on this point briefly.

If we consider the K\"ahler form in (\ref{2-10}),
the potential $W$ can be absorbed into the ambiguities in
the choice of the K\"ahler potential since, as explicitly shown in (\ref{3-9}),
$W$ is given in a form of separation of holomorphic and antiholomorphic parts.
On the other hand, in terms of $(a_1 , a_2 )$ the potential
$K_0$ can be expressed as
\beq
    K_0  \, = \, \frac{\pi l}{ \tau_2 } a \ba \, = \,   - \frac{\pi l}{4 \tau_2} \left[
    a_1^2 + \left( \frac{a_2}{\tau_2} \right)^2
    \right] \, .
    \label{3-15}
\eeq
Thus it is given in a form of $u(a_1) + v (a_2)$ as well. This implies
that $W( a_1 , a_2 ) = i \frac{\pi l}{2 \tau_2} a_1 a_2$ in (\ref{3-9})
is considered to be a counter part of $K_0 = \frac{\pi l}{ \tau_2 } a \ba$
in the $( a_1 , a_2 )$ representation.
As mentioned elsewhere, $( a_1 , a_2 )$ are canonical coordinates
of the phase space of interest whose symplectic form is given by (\ref{3-7}).
Thus, classically, we can describe physical observables
in terms of $( a_1 , a_2 )$.
As far as the K\"ahler potential and the symplectic
potential are concerned, we can then define these in terms of $( a_1 , a_2 )$.
{\it Quantum theoretically, however, we need to impose the polarization condition (\ref{2-7})
on the zero-mode wave function.}
We therefore require that $f(a)$ as defined in (\ref{3-12}, \ref{3-14})
should be a holomorphic function of $a$.

\vskip 0.5cm \noindent
\underline{Gauge transformations induced by doubly periodic translations}

As mentioned earlier, holomorphic functions on the torus in general
obey the double periodicity condition.
Thus we can naturally assume $f ( a ) = f (a + m + in \tau_2 )$.
In what follows, we show how this assumption can be understood by
imposing an invariance on the wave function
$F[ a, \ba ] = e^{ - \frac{W}{2} } f(a) $ under a gauge transformation
of $\A^{( \tau_2 )}$ in (\ref{3-8}) where the gauge transformation
is induced by the doubly periodic
translations $a \rightarrow a + m + i n \tau_2$.

Under $a \rightarrow a + m + i n \tau_2$, $a_1$ and $a_2$ vary as
\beq
    a_1 \, \rightarrow \, a_1 - 2 i n \tau_2 \, , ~~~
    a_2 \, \rightarrow \, a_2 + 2 i m \tau_2  \, .
    \label{3-16}
\eeq
The symplectic potential (\ref{3-8}) transforms as
\beq
    \A^{(\tau_2 )} \, \rightarrow \, \A^{(\tau_2 )} \, + \, d \La_{m,n}
    \label{3-17}
\eeq
where
\beq
    \La_{m,n} \, = \, - i \frac{\pi l}{2 \tau_2 } \, ( m a_1 - n a_2 ) \, .
    \label{3-18}
\eeq
The gauge invariance of the wave function $F[ a, \ba ] = e^{ - \frac{W}{2} } f(a) $
is then realized by the relation
\beq
    e^{i \La_{m,n}} F[ a, \ba ] \, = \, F[a + m + i n \tau_2 , \ba + m - i n \tau_2] \, .
    \label{3-19}
\eeq
Explicit forms of the left and right-hand sides are given by
\beqar
    {\rm (l.h.s)} &=&
    \exp \left[ \frac{\pi l}{ 2 \tau_2} ( m a_1 - n a_2 ) - i \frac{\pi l}{4 \tau_2^2} a_1 a_2 \right]
    f(a) \, ,
    \label{3-20} \\
    {\rm (r.h.s)} &=&
    \exp \left[ -i \frac{\pi l}{4 \tau_2^2} ( a_1 - 2 in \tau_2)( a_2 + 2i m \tau_2 ) \right]
    f(a + m + in \tau_2 ) \, .
    \label{3-21}
\eeqar
The gauge invariance condition (\ref{3-19}) thus leads to the relation
\beq
    e^{i \pi l mn}  f(a) \, = \, f(a + m + i n \tau_2 ) \, .
    \label{3-22}
\eeq
The consequence of the gauge invariance (\ref{3-19}) is the following;
the holomorphic zero-mode wave function $f (a)$ is
invariant under $a \rightarrow a + m + i n \tau_2$,
given that the level number $l$ is quantized by even integers, {\it i.e.},
\beq
    l \, \in  \, 2 \Z \, .
    \label{3-23}
\eeq
This level-number quantization condition has been known for the
abelian CS theory on the torus \cite{Bos:1989wa,NairBook}.
The relation (\ref{3-22}) was first reported in
\cite{Abe:2008wn} and further developed in \cite{Abe:2010sa}
but in these papers the ambiguities in the choice
of the K\"{a}hler potential has not been carefully treated.
We here argue that the use of the particular wave function
$F[ a, \ba ]$ in (\ref{3-13}) appropriately leads to
the relation (\ref{3-22}).

Lastly, we would like to comment on the irreducibility of
the symplectic potential
$\A^{( \tau_2 )}  = - \frac{ \pi l}{4 \tau_2^2 } \left( a_1 da_2 + a_1 d a_2 \right)$.
in (\ref{3-8}).
As mentioned before, $a_1$ and $a_2$ can serve as
the canonical coordinates of a physical system. So it may be possible
to impose a ``polarization'' condition  on this $\A^{( \tau_2 )}$.
The doubly periodic translations, however, involve the both
winding numbers $(m ,n )$, corresponding to the variations of
$a_2$ and $a_1$, respectively, as shown in (\ref{3-16}).
Thus, in order to express the gauge transformation (\ref{3-17})
such that these winding numbers are impartially treated,
we need to define $\A^{( \tau_2 )}$ in terms of both $a_1$ and $a_2$ explicitly.
In this sense (\ref{3-8}) provides an irreducible representation
for the symplectic potential whose gauge transformations are induced by
the doubly periodic translations.

\section{Holomorphic wave functions as  modular forms of weight 2}

In this section we briefly present the main results in the previous paper \cite{Abe:2017mf}.
The upshot of the previous paper is that under the modular
$S$- and $T$-transformations a holomorphic zero-mode wave function $f(a)$  in abelian
Chern-Simons theory on the torus varies as
\beqar
    S: \, f \left( - \frac{1}{a} \right) &=& a^2 f(a) \, ,
    \label{4-1} \\
    T: \, f ( a  + 1 ) &=& f(a) \, ,
    \label{4-2}
\eeqar
given that $f(a)$ is quantum theoretically characterized by the operative relation
\beq
    \frac{\d}{\d a } f (a) \, = \, \frac{\pi l}{ \tau_2} \ba \, f(a)
    \label{4-3}
\eeq
and the inner product of the zero-mode wave functions
\beq
    \< \Psi | \Psi^\prime \> \, = \,
    \int d\mu(a,\ba) \, e^{-K(a,\ba)} \, \overline{f(a)} f^\prime (a)
    \label{4-4}
\eeq
where the zero-mode wave function is defined as $\Psi [a ,\ba ] = e^{- \frac{K(a, \ba)}{2}} f(a)$.
The operative relation (\ref{4-2}) is guaranteed as long as
the K\"ahler potential is given in the form of
$K(a, \ba) = K_0 + u(a) + v (\ba )$ where $K_0 = \frac{\pi l}{\tau_2 } a \ba$.
Notice that our choice (\ref{3-10}) in the previous section falls within this form.

The modular transformations of our interest are generated by
\beqar
    S: \, (a , \tau_2 ) & \rightarrow & \left( - \frac{1}{a} , \frac{\tau_2}{|a|^4} \right) \, ,
    \label{4-5} \\
    T: \, (a , \tau_2 )  & \rightarrow & ( a  + 1 , \tau_2 ) \, .
    \label{4-6}
\eeqar
These imply that the quantity $\frac{|da|^2}{\tau_2}$,
or the number density of the zero modes per unit area,
is preserved under the modular transformations of $a$.
Note that under the $S$-transformation the area element $da d \ba$ changes as
$|da|^2 \rightarrow \frac{|da|^2 }{|a|^4}$.
Thus from (\ref{4-5}) we find that the number density
$\frac{|da|^2}{\tau_2}$ is preserved under the $S$-transformation.
The modular $T$-invariance of the number density is obvious.

\vskip 0.5cm \noindent
\underline{Basics of modular forms}

We review some basic facts on the modular forms.
In general, the modular form $f(z)$ of weight $k$ is defined by
\beq
    f \left( \frac{ \al z + \bt }{\ga z + \del } \right) \, = \, ( \ga z + \del )^k  f(z)
    \label{4-7}
\eeq
where $\al$, $\bt$, $\ga$, $\del$ are matrix elements of the modular group
\beq
    SL( 2, \Z ) \, = \, \left\{ \left.
    \left(
      \begin{array}{cc}
        \al & \bt \\
        \ga & \del \\
      \end{array}
    \right) \right| \, \al, \bt, \ga, \del \in \Z , ~ \al \del - \bt \del = 1
    \right\} \, := \, \Ga  \, .
    \label{4-8}
\eeq
The modular forms are defined on the upper-half plane
$\H = \{ z \in \C \, | \, \im \, z > 0 \}$.
Accordingly, to be rigorous, the modular group is defined as
$PSL( 2 , \Z ) := SL( 2 , \Z ) / \{ \pm I \}$, with $I$ the identity matrix.
The fundamental domain ${\cal F}$ for the action of
$SL (2 , \Z)$ generators on $\H$ is given by
\beq
    {\cal F} \, = \, \left\{ z \in \C \left| \,  \im \, z > 0 , ~
    |z| \ge 1, ~ |\re \, z | \le \hf \right. \right\} \, .
    \label{4-9}
\eeq

The generators of the modular group is given by
$\left( \!
  \begin{array}{cc}
    1 & 1 \\
    0 & 1 \\
  \end{array}
\! \right)$ and $
\left( \!
  \begin{array}{cc}
    0 & -1 \\
    1 & 0 \\
  \end{array}
\! \right)$.
The definition of the modular form in (\ref{4-7}) is then obtained
from the conditions
\beqar
    f ( z + 1 ) &=& f(z) \, ,
    \label{4-10} \\
    f \left( - \frac{1}{z} \right) &=& z^k f( z) \, .
    \label{4-11}
\eeqar
The first condition simply means that
$f(z)$ can be expressed in a form of the Fourier expansion
\beq
    f(z) \, = \, \sum_{n = 0}^{\infty} a_n \, q^{n}
    \label{4-12}
\eeq
where $q = e^{i 2 \pi z }$ and $a_n$ is the Fourier coefficient.
If $a_0 = 0$, the modular form $f(z)$ is called the {\it cusp form}.
The vector space formed by the cusp forms of weight $k$ is denoted
by $S_k ( \Ga )$, {\it i.e.},
\beq
    S_k ( \Ga ) \, := \, \left\{
    f: \H \rightarrow \C \left| \,
    f \left( - \frac{1}{z} \right) = z^k f( z) , \,
    f(z) =  \sum_{n = 1}^{\infty} a_n \, q^{n}
    \right.
    \right\} .
    \label{4-13}
\eeq

Let $f(z) , \, g(z) \in S_k (\Ga )$, then the Petersson inner product is defined as
\beq
    \< f , g \> \, = \, \frac{1}{{\rm vol}  {\cal F}  } \int_{\cal F}
    f(z) \overline{ g (z) } \, y^k \, \frac{dx  dy }{y^2}
    \label{4-14}
\eeq
where $z = x + i y$. Notice that this inner product represents
a manifestly modular invariant integral.

\vskip 0.5cm \noindent

From (\ref{4-1}, \ref{4-2}) and (\ref{4-10}, \ref{4-11}) we can quantum theoretically identify
$f(a)$ as a modular form of weight 2.
Consequently, we may define the zero-mode variable on
the fundamental domain ${\cal F}$ in (\ref{4-7}).
The inner product of the zero-mode wave functions (\ref{2-22}) can then be rewritten as
\beqar
    \< \Psi | \Psi^\prime \> & = &
    \int d\mu(a,\ba) \, e^{-K(a,\ba)} \, \overline{f(a)} f^\prime (a)
    \nonumber \\
    & \rightarrow &
    \frac{1}{{\rm vol}  {\cal F}  } \int_{\cal F}
    d (\re \, a) d (\im \, a)
    \, e^{-K(a,\ba)} \, \overline{f(a)} f^\prime (a)
    \label{4-15}
\eeqar
where we express the the integral measure $d \mu ( a, \ba )$
as $d (\re \, a) d (\im \, a) $ in the second line.
This inner product can be considered as a quantum version of
the Petersson inner product (\ref{4-14}) for the modular forms of weight 2.

\section{Physical views on the Hecke operators}

\vskip 0.5cm \noindent
\underline{Basics on the Hecke operators}

We now introduce basic ideas of Hecke operators acting on the modular
forms of weight $k$, following mathematical textbooks, {\it e.g.},
\cite{Koblitz:1993bk,Ono:2004bk,Kurokawa:2005bk,Stein:2007bk,Lozano-Robledo:2011bk}.
In order to define the Hecke operators it is useful
to view the modular forms as functions on complex lattices.
Let $f (L)$ be a function on a lattice $L$ in $\C$. Then a Hecke operator $T_m$
acting on $f (L)$ is defined as
\beq
    T_m \, f ( L ) \, = \,  \sum_{[ L: L^\prime ] = m } f ( L^\prime )
    \label{5-1}
\eeq
where the sum is taken over all sublattices $L^\prime \subset L$ of index $m$.
Note that a sublattice $L^\prime \subset L$ has index $m$ if
the quotient $L/ L^\prime$ has order dividing $m$ so that $m L \subset L^\prime
\subset L$ and
\beq
    L^\prime / m L \, \subset \, L / m L \, = \, ( \Z / m \Z )^2 \, .
    \label{5-2}
\eeq
This means that the sublattices of index $m$ correspond to
the subgroups of order $m$ in $( \Z / m \Z )^2$.
If $m$ is a prime number $p$, there are $p + 1$ such subgroups,
since the number of the subgroups corresponds to
the number of nonzero vectors in $\F_p $ modulo scalar equivalence,
and there are $( p^2 -1 ) / (p - 1 ) = p+1$ such vectors \cite{Stein:2007bk}.

It is known that there is a one-to-one correspondence
between sublattices $L^\prime \subset L$ of index $m$
and matrices
$
\left(
  \begin{array}{cc}
    \al & \bt \\
    0 & \del \\
  \end{array}
\right)
$
with $\al, \bt , \del \in \Z$, $ \al \del = m$ and $ 0 \le \bt \le \del - 1$;
for details and a proof, see \cite{Stein:2007bk}.
The choice of the index-$m$ sublattices is then reduced to
that of the matrix elements in the above matrix with determinant $m$.
By use of this fact, one finds that the Hecke operator acting
on $f (z) \in S_k (\Ga )$ can be defined as
\beqar
    T_m f (z) &=& m^{k-1} \sum_{\al \del = m}
    \sum_{\bt = 0}^{\del -1} \del^{-k} f \left( \frac{ \al z + \bt }{\del} \right)
    \nonumber \\
    &=&
    \sum_{n=1}^{\infty} \left( \sum_{\al | ( m , n)} \al^{k-1} \, a_{\frac{m n}{\al^2}}
    \right) q^n
    \label{5-3}
\eeqar
where the Fourier expansion of $f(z) = \sum_{n \ge 1} a_n q^n$ is given by (\ref{4-13}) with
$q = e^{i 2\pi z}$, and $( m , n)$ denotes $gcd ( m , n)$.
This means that the Hecke operator $T_m$ preserves
the space of modular forms of a given weight,
$T_m : S_k ( \Ga ) \rightarrow S_k ( \Ga )$.
When the index $m$ is prime $m = p$, the above expression simplifies as
\beq
    T_p \, f (z) \, = \, \sum_{n = 1}^{ \infty } \left(
    a_{pn} + p^{k-1} \, a_{\frac{n}{p} }
    \right) \, q^n \, .
    \label{5-4}
\eeq

The Hecke operator forms an abelian algebra
\beq
    T_m \, T_n \, = \, T_n \, T_m \, = \,
    \sum_{\al | ( m, n)} \al^{k-1} \, T_{\frac{mn}{\al^2}}
    \label{5-5}
\eeq
From this relation we find $T_m T_n = T_n T_m = T_{mn}$ if $(m, n ) = 1$.
Suppose $f(z) \in S_k ( \Ga )$ is a simultaneous eigenfunction of the
Hecke operators $T_m$  for all $m= 1,2, \cdots$, {\it i.e.},
if there exists a set of  eigenvalues $\la_m$ such that
\beq
    T_m \, f(z) \, = \, \la_m \, f(z) \, ,
    \label{5-6}
\eeq
then the following form of a Dirichlet series can be
expressed as an Euler product:
\beqar
    L ( s, f ) &=&
    \sum_{n \ge 1} \la_n \, n^{-s}
    \nonumber \\
    &=& \prod_{p: \, prime}
    \frac{1}{1 \, - \, \la_p  \, p^{-s} \, + \,  p^{k -1 -2s } }
    \label{5-7}
\eeqar
where $s \in \C$. This function of $s$
is called the {\it $L$-function} of the modular form $f(z)$.

As described in (\ref{4-14}), the vector space of the cusp forms $f(z), g(z) \in S_k ( \Ga )$
has the Petersson inner product $\< f , g \> $.
It is well known that in terms of this inner product the Hecke operator $T_m$ is a hermitian
operator; see \cite{Koblitz:1993bk} for details.
In other words, we have $\< T_m  f , g \> = \< f , T_m  g \> $ and,
accordingly, the eigenvalue is a real constant $\la_m \in \R$.
In fact, from (\ref{5-3}) and (\ref{5-6}) we can show that
\beq
    \la_m \, = \, a_m
    \label{5-8}
\eeq
with the normalization of $f(z)$ by $a_1 = 1$.
This can easily be seen by expanding $T_m f(z) = \la_m f(z)$ as
$T_m f(z) = \sum_{n \ge 1} b_n q^n$ and reading off the coefficient
$b_1$ from (\ref{5-3}), which leas to $b_1 = a_m = \la_m a_1$.

\vskip 0.5cm \noindent
\underline{Level $N$ congruence subgroup $\Ga_0 (N)$ of $\Ga$}

A level $N$ {\it congruence subgroup} $\Ga_0$ of the modular group $\Ga$
is defined as
\beq
    \Ga_0 (N) \, := \, \left\{ \left.
    \left(
      \begin{array}{cc}
        \al & \bt \\
        \ga & \del \\
      \end{array}
    \right) \in SL( 2, \Z ) \right| \ga \equiv 0 ~ ({\rm mod}~ N)
    \right\}
    \label{5-9}
\eeq
There exist modular forms corresponding to $\Ga_0 (N )$, {\it i.e.},
those that satisfy the definition (\ref{4-7}) by the matrix elements in (\ref{5-9}).
In terms of such modular forms a congruence subgroup of
the vector space $S_k ( \Ga )$ in (\ref{4-13}) is similarly defined and is conventionally
denoted by $S_k ( \Ga_0 (N) )$.
Obviously, $\Ga_0 (N)$ and $S_k ( \Ga_0 (N) )$ reduce to $\Ga$ and $S_k ( \Ga )$,
respectively, at $N= 1$.

Introducing a Dirichlet character $\chi$ modulo $N$, we can define
a relevant cusp form $f (z) \in S_k ( \Ga_0 (N) , \chi )$ by
\beq
    f \left( \frac{ \al z + \bt }{\ga z + \del } \right) \, = \,
    \chi ( \del ) \, ( \ga z + \del )^k  f(z)
    \label{5-10}
\eeq
where $\al$, $\bt$, $\ga$, $\del$ are matrix elements of $\Ga_0 (N)$.
In analogy to (\ref{5-3}), the action of the Hecke operator $T_m$ on
$f (z) = \sum_{n \ge 1} a_n q^n \, \in S_k ( \Ga_0 (N) , \chi )$
can be defined by \cite{Ono:2004bk}
\beq
    T_m f (z) \, = \,
    \sum_{n=1}^{\infty} \left( \sum_{\al | ( m , n)}
    \chi ( \al ) \al^{k-1} \, a_{\frac{m n}{\al^2}}
    \right) q^n
    \label{5-11}
\eeq
Suppose that $f(z) \in  S_k ( \Ga_0 (N) , \chi )$ is
is a simultaneous eigenfunction of the above
Hecke operators $T_m$  for all $m= 1,2, \cdots$, that is,
we have $T_m f(z) = \la_m f(z)$.
Since $T_m f(z) \in S_k ( \Ga_0 (N) , \chi )$  as well,
we can expand it as $T_m f(z) = \sum_{n \ge 1} b_n q^n$.
From (\ref{5-11}) and the identity of the Dirichlet character $\chi (1) = 1$,
we then find
\beq
    b_1 \, = \, a_m \, = \, \la_m \, a_1 \, = \la_m
    \label{5-12}
\eeq
with the normalization $a_1 = 1$.
In this case the corresponding $L$-function can be expressed as
\beqar
    L ( s, f ) & = &
    \sum_{m \ge 1} \la_m m^{-s}
    \nonumber \\
    &=& \prod_{p \mid N} \frac{1}{1 - a_p p^{-s}} \,
    \prod_{p \nmid N} \frac{1}{1 - a_p p^{-s} + \chi(p) p^{k-1-2s}}
    \label{5-13}
\eeqar
where, as in (\ref{5-7}), $p$ denotes the prime numbers.
Notice that the Dirichlet character $\chi(p)$ of modulo $N$ vanishes
whenever $p \mid N$; thus we can define the above $L$-function
simply as $L (s ,f ) = \prod_{p} ( 1 - a_p p^{-s} + \chi (p) p^{k-1-2s} )^{-1}$
without splitting into factors of $p \mid N$ and $p \nmid N$.

\vskip 0.5cm \noindent
\underline{Hecke operators acting on the holomorphic wave function $f(a)$}

In the previous sections we have argued that the
holomorphic zero-mode wave function $f(a)$ in abelian Chern-Simons
theory on the torus can be considered as a quantum version of
a modular from of weight 2.
In the following we think of how the Hecke operator arises as an action to $f(a)$.
To begin with, we recall that the gauge invariance condition
(\ref{3-19}) for the zero-mode wave function leads to the relation
$e^{i \pi l mn}  f(a)  =  f(a + m + i n \tau_2 )$ in (\ref{3-22}).
For $l \in 2 \Z$, $f(a)$ satisfies the doubly periodic condition
and we can identify $f(a)$ as a holomorphic function defined on a complex lattice.
For $l$ being an odd integer, say $l=1$, we have
\beq
    f(a + m + i n \tau_2 )\, = \, (-1 )^{mn} f(a)  \, .
    \label{5-14}
\eeq
Regarding the factor $(-1)^{mn}$ as a ``phase'' factor, we can also
consider $f(a)$ as a function on the complex lattice.
Thus the above-mentioned definition of the Hecke operator
(\ref{5-1})-(\ref{5-5}) applies to $f(a)$ as well.
One of the peculiarities in $f(a)$, distinguished from classical functions
on the torus, is given by the relation (\ref{5-14}).
In order to investigate quantum properties of $f(a)$
we fix $l$ at $l = 1$, while keeping $\tau_2$ finite, in the following.

From (\ref{5-1}) we see that the Hecke operator $T_M$ acting
on a function $f(L)$ on a complex lattice $L$ is defined
as the sum over sublattices $L^\prime \subset L$ of index $M$.
In terms of the holomorphic wave function $f(a)$ this
can be expressed as
\beqar
    T_M f(L) &=& \sum_{[ L , L^\prime ] = M} f(L^\prime )
    \nonumber \\
    \longrightarrow ~~
    T_M f (a)  & = & \sum_{ m \in \F_M } \sum_{ n \in \F_M } f( a + m + i n \tau_2 )
    \nonumber \\
    &=& \sum_{m ,n \in \F_M} (-1 )^{mn} f(a)
    \label{5-15}
\eeqar
where we use (\ref{5-2}), (\ref{5-14}) and $\F_M = \Z / M \Z$.
Naively, this means that $f(a)$ is an eigenfunction of the Hecke operator,
$T_M f(a) = \la_M f(a)$ where the eigenvalue is given by
\beq
    \la_M \, = \, \sum_{m ,n \in \F_M} (-1 )^{mn}  \, .
    \label{5-16}
\eeq
As reviewed earlier, the eigenvalue corresponds to the Fourier coefficient
of $f(a)$ and defines the $L$-function of interest. Thus it is intriguing
if we can compute this value.
Although the expression (\ref{5-16}) suggests that $\la_M$ are integers,
this expression is rather intuitive and not well-defined compared to that of (\ref{5-3}).
For example, the sum over $m, n \in \F_M$
means a change of fields for $m, n$ since these are initially defined as integers,
corresponding to the winding numbers along $\al$ and $\bt$ cycles
on the torus, respectively.
Within the interpretation of the doubly periodic translations $a \rightarrow a + m + i n \tau_2$
as a combination of modular transformations,
this implies that we change the  matrix elements of the modular
group from integer to finite field,  {\it i.e.}, $SL ( 2 , \Z ) \rightarrow SL ( 2, \Z / M \Z )$.
Thus, the notion of the level for $f(a)$ naturally arises from
an interpretation of (\ref{5-16}).
In other words, in order to compute the value of (\ref{5-16})
it would be suitable to consider $f (a)$ as level $M$ cusp forms
of weight 2, $f(a) \in S_2 ( \Ga_0 (M) )$.

\vskip 0.5cm \noindent
\underline{A speculative connection to the Legendre symbol}

As a digression, we now briefly discuss a speculative idea
on the interpretation of the factor $( -1 )^{mn}$.
The scale invariance of the zero-mode coordinate in (\ref{2-9})
suggests an implicit condition $gcd( m , n) = 1$.
One of the simplest choices would be $(m , n) = ( p, q )$
where $p, q$ are (odd) prime numbers.
Such a choice reminds us of mathematical analogies between
primes and knots \cite{Morishita:2009gt}.
Previously in \cite{Abe:2010sa}, we argue that the
factor $(-1)^{mn}$ acting on $f(a)$ can be interpreted
as $( -1 )^{lk ( \al ,\bt )}$ where $lk ( \al ,\bt )$ denotes
a liking number of the $\al$ and $\bt$ cycles along the torus.
With the choice of $(m , n) = ( p, q )$,
the linking number becomes
$( -1 )^{lk ( p , q )}$. Then, by use of mathematical analogies between linking numbers
and Legendre symbols \cite{Morishita:2009gt}, we have
\beq
    (-1)^{lk ( p , q )} \, \longleftrightarrow \, \la_{p} ( q)
    \label{5-17}
\eeq
where $\la_{p} ( q)$ denotes the Legendre symbol, with $p$ and $q$ being odd primes.
In terms of the conventional notation
this can also be expressed as
\beq
    \la_{p} ( q) \, = \, \left( \frac{q}{p} \right)
    \, = \, \left( \frac{p^*}{q} \right) \, = \, \la_{q} ( p^*)
    \label{5-18}
\eeq
where $p^* = (-1)^{\frac{p-1}{2}} p$ and we have used the reciprocity law
of the Legendre symbol $\left( \frac{q}{p} \right)  =  \left( \frac{p^*}{q} \right)$.

Since the Legendre symbol gives a map $\la_p ( q ) : \F_p \rightarrow \C$
it is natural to consider an action of it to the holomorphic wave function
$f(a)$ in terms of its Fourier transform:
\beq
    \widehat{ \la}_{p}  \, := \, \sum_{q \in \F_{p}} \la_{p} ( q)
    \, e^{i \frac{2 \pi}{p} q } \, = \, \sqrt{p^*} 
    \label{5-19}
\eeq
which is known as the Gauss sum.
Once we choose and fix the pair $(p, q)$ and bear in mind the above analogies,
we may speculate that 
the action of $\sum_{m, n}(-1)^{mn}$ on $f(a)$ defined in the $\C$-space
would be described by the Gauss sum or a normalized value of it.
To make this statement a bit clearer, let us compute the eigenvalues $\la_N$ for the
level $N$ cusp forms of weight 2, $f(a) \in S_2 ( \Ga_0 (N) )$
with $N$ being odd primes.
According to \cite{LMFDB}, such cusp forms become dimension 1
only for $N = 11, 17, 19$ and in each case the coefficient
$\la_N$ for the corresponding $L$-function is given by
$\la_N = 1 / \sqrt{N}$. 
Note that these values can be read off from a list of analytically normalized
$L$-functions \cite{LMFDB}.
Notice also that the corresponding coefficient $a_N$ of the $q$-expansion
of the cusp form $f(a) \in S_2 ( \Ga_0 (N) )$ is given by $a_N = 1$ 
for $N= 11, 17, 19$. Thus the factor of $1 / \sqrt{N}$ may be interpreted
as an overall normalization factor but other coefficients 
$\la_m$ ($m \ne N$) of the $L$-function as a Dirichlet series 
are expressed as $\la_m = \Z / \sqrt{m}$ \cite{LMFDB}.
Thus we may not consider $1 / \sqrt{N}$ as an overall normalization
factor for the $L$-functions of $f(a) \in S_2 ( \Ga_0 (N) )$.
Following the above discussion, we find that these values $(\la_N = 1 / \sqrt{N})$
may be interpreted as a normalized Gauss sum $\sqrt{N}/ N = 1 / \sqrt{N}$
with the choice of $p = N^*$.
We have tried to develop these ideas to understand other coefficients 
$\la_m$ ($m \ne N=11,17,19$) of the $L$-functions
for $f(a) \in S_2 ( \Ga_0 (N) )$  but, at the present,
we do not have any satisfactory explanations for these values.

\vskip 0.5cm \noindent
\underline{Physical interpretation of the Hecke operator}

In this section, we have considered how the Hecke operators act
on the holomorphic zero-mode wave function $f(a)$ in
abelian Chern-Simons theory on the torus by use of
the relation (\ref{5-14}).
We first introduce the formal definition (\ref{5-1}) of the Hecke operators
acting on a function on a complex lattice. Applying this definition
to $f(a)$, we find the expression (\ref{5-15}). This naturally gives rise to
the notion of the level for $f(a)$ as the modular form.
As mentioned in (\ref{3-19}), the relation (\ref{5-14}) arises from
invariance under gauge transformations induced by
the doubly periodic translations $a \rightarrow a + m + i n \tau_2$ ($m ,n \in \Z$)
of the zero-mode variable $a \in \C$.
Therefore, from a gauge theoretic perspective,
it is straightforward that $f(a)$ is an eigenform of the Hecke operator. 
{\it In this context we can interpret the action
of the Hecke operator on $f(a)$ as a sum of the possible gauge transformations
of $f(a)$ induced by the doubly periodic translations.}
If $f(a)$ is an eigenform of the Hecke operator, then there automatically exits
a corresponding $L$-function for $f(a)$, with the Dirichlet characters ($\la_m$ in (\ref{5-13}))
given by the eigenvalues of $f(a)$.
Thus it is intriguing if we can understand the eigenvalues (\ref{5-16})
from a physical perspective. We briefly sketch that part of such values may
be computed by use of mathematical analogies between
linking numbers and Legendre symbols \cite{Morishita:2009gt}.

\section{Conclusion}

In the previous paper \cite{Abe:2017mf} we show that
the holomorphic zero-mode wave function $f(a)$ in abelian
Chern-Simons theory on the torus can be considered
as a quantum version of a modular form of weight 2.
Motivated by this result, in this paper we consider
how a Hecke operator acts on $f(a)$, in hope of
obtaining physical interpretations of the Hecke operators
and corresponding $L$-functions in number theory.
The Hecke operators are formally defined
as a sum of sublattices on which modular forms in general are defined.
The modular forms can be considered as holomorphic functions on a complex lattice
or a (complex) torus. Such functions generally satisfy the doubly periodic conditions.

In the first half of this paper we review that the
holomorphic wave function $f(a)$ satisfies the doubly periodic
condition $f(a) = f ( a +  m + i n \tau_2 )$ (with $m, n \in \Z$ and $\tau_2 > 0$)
when the level number $l$ of the Chern-Simons theory is even.
To be more precise, we show that the gauge invariance condition
(\ref{3-19}) for the zero-mode wave function leads to the relation
$e^{i \pi l mn}  f(a)  =  f(a + m + i n \tau_2 )$ in (\ref{3-22}).
We can then interpret $f(a)$ as a holomorphic function defined on
the complex lattice as well, with the factor of  $e^{i \pi l mn}$ representing
quantum effects.
Nontrivial quantum effects are given by $l$ being an odd integer.

In the latter half of the paper, we consider an action of the
Hecke operator acting on $f(a)$ with $l = 1$.
From the formal definition of the Hecke operator (\ref{5-1})
we argue that the action of it on $f(a)$ can be described as a sum of
the gauge transformations of $f(a)$ induced by $a \rightarrow a + m + i n \tau_2$.
In order to make sense of the resultant expression (\ref{5-15})
we also argue that the notion of the level naturally arises for $f(a)$ as a modular form.
Our interpretation of the Hecke operator, {\it i.e.}, as
a sum over possible gauge transformations of $f(a)$, automatically
indicates that $f(a)$ is an eigenform of the Hecke operator.
This, on the other hand, guarantees
the existence of the corresponding $L$-function for $f (a)$
where $f(a)$ can be seen as a level $N$ cusp form of weight 2, $f (a) \in S_2 (\Ga_0 (N) )$.

We also present a speculative idea that eigenvalues
$\la_N$ (with $N$ being an odd prime) for such $f (a) \in S_2 (\Ga_0 (N) )$
may be computed by use of mathematical
analogies between linking numbers and Legendre symbols \cite{Morishita:2009gt}.
According to \cite{LMFDB}, we have $\dim [ S_2 (\Ga_0 (N) ) ] = 1$ for $N= 11,17, 19$
and in these particular cases $\la_N$ as the Dirichlet characters of
of the corresponding $L$-functions are given by $\la_N = 1 / \sqrt{N}$.
We observe that these values may be interpreted as normalized
versions of the Gauss sum $\widehat{ \la}_{N^*} = \sqrt{N}$ which can be seen as
a Fourier transform of the corresponding Legendre symbol.
Unfortunately, these ideas are still at a speculative stage
but, hopefully, would shed some new light on physical approaches to problems
in number theory.


\vskip .3in

\end{document}